\documentstyle[prl,aps,preprint,epsf]{revtex}
\begin{document}
\draft
\title{DC-electric-field-induced and low-frequency electromodulation second-harmonic 
generation 
spectroscopy of Si(001)-SiO$_2$ interfaces}
\author{ O. A. Aktsipetrov\footnote{E-mail address: aktsip@astral.ilc.msu.su, Web: 
http://kali.ilc.msu.su}, A. A. Fedyanin, A. V. Melnikov, E. D. Mishina, and A. N. 
Rubtsov }
\address{Department of Physics, Moscow State University, Moscow 119899, Russia}
\author{M. H. Anderson, P. T. Wilson, M. ter Beek, X. F. Hu, J. I. Dadap\footnote{ 
Present 
address: Department of Physics and Department of Chemistry, Columbia University, 
New York, NY 10027-6902}, and M. C. Downer}
\address{Department of Physics, The University of Texas at Austin, Austin, TX 78712}

\date{\today}
\maketitle

\begin{abstract}
The mechanism of DC-Electric-Field-Induced Second-Harmonic (EFISH) generation at 
weakly 
nonlinear buried Si(001)-SiO$_2$ interfaces is studied experimentally in planar 
Si(001)-SiO$_2$-Cr MOS structures by optical second-harmonic generation (SHG) spectroscopy with a 
tunable 
Ti:sapphire femtosecond laser. The spectral dependence of the EFISH contribution near 
the direct 
two-photon $E_1$ transition of silicon is extracted. A systematic phenomenological 
model of the 
EFISH phenomenon, including a detailed description of the space charge region (SCR) at 
the 
semiconductor-dielectric interface in accumulation, depletion, and inversion regimes, has 
been 
developed. The influence of surface quantization effects, interface states, charge traps in 
the oxide 
layer, doping concentration and oxide thickness on nonlocal screening of the DC-electric 
field and 
on breaking of inversion symmetry in the SCR is considered. The model describes EFISH 
generation in the SCR using a Green function formalism which takes into account all 
retardation 
and absorption effects of the fundamental and second harmonic (SH) waves, optical 
interference 
between field-dependent and field-independent contributions to the SH field and multiple 
reflection 
interference in the SiO$_2$ layer. Good agreement between the phenomenological 
model and our 
recent and new EFISH spectroscopic results is demonstrated. Finally, low-frequency 
electromodulated EFISH is demonstrated as a useful differential spectroscopic technique 
for 
studies of the Si-SiO$_2$ interface in silicon-based MOS structures.

\end{abstract}

\vspace{5mm}
\noindent
PACS numbers: 42.65.Ky, 73.65.Qv, 68.35.-p

\newpage

\section{Introduction}
\par
Optical Second Harmonic Generation (SHG) has been one of the most intensively 
studied 
phenomena in surface and interface optics\cite{1,2,3} for the last decade. The interest in 
SHG 
stems 
from its unique sensitivity to the structural and electronic properties of surfaces and 
interfaces of 
centrosymmetric media. This unusually high surface/interface-sensitivity comes about 
because, in 
the electric dipole approximation, SHG is forbidden in the bulk of materials with 
inversion 
symmetry\cite{4,5}, but allowed at interfaces, where inversion symmetry is broken by 
the 
discontinuity 
of crystalline structure. Related nonlinear sources of SHG are localized in a thin (several 
nanometers thick) surface or interface layer. In semiconductors, inversion symmetry is 
also broken 
by the DC-electric Field (DCF) in the subsurface Space Charge Region (SCR), which is 
created by 
initial band bending and/or external bias application. The lack of inversion symmetry in 
the SCR 
results in DC-Electric-Field Induced Second-Harmonic (EFISH) generation, which 
manifests itself 
through electromodulation of the SHG intensity. Thus, all important properties of 
surfaces, buried 
interfaces and subsurface layers - their charge\cite{6,7,8}, electronic surface state 
density
\cite{9,10,11}, 
roughness (morphology)\cite{12,13}, adsorption (adatom and admolecule surface 
density)\cite{14,15,16,17}, 
initial band bending\cite{18,19,20}, {\it etc.} - can, in principle, be determined by means 
of the SHG 
probe.
\par
The technological importance of Si(001)-SiO$_2$ interfaces stems from their ubiquitous 
presence 
in Metal-Oxide-Semiconductor (MOS) structures and MOS Field Effect Transistors 
(MOSFET). 
EFISH generation provides a promising noninvasive, {\it in situ}  technique for characterizing 
interfacial imperfections and charge defects at the Si(001)-SiO$_2$ interface. Moreover, 
the 
relative simplicity of the description of the SHG response from the Si(001) face, 
originating from 
the small number of tensor components of the interface quadratic susceptibility and the 
rotationally 
isotropic interfacial SHG response, makes Si(001)-SiO$_2$ interface among the most 
important for 
investigation of fundamental aspects of the EFISH phenomenon.
\par
The 1967 discovery of EFISH generation by Bloembergen and co-workers\cite{21} at Si- 
electrolyte and 
Ag-electrolyte interfaces in electrochemical cells remained largely unnoticed for a 
number of 
years. The 1981 discovery of surface-enhanced SHG by Shen and co-workers\cite{22} 
rejuvenated 
interest in this effect. Surface-enhanced EFISH generation at a silver-electrolyte interface 
was 
observed shortly afterward\cite{23}. Since 1984 EFISH has been systematically studied 
at 
Si(111)-
electrolyte interfaces\cite{24,25,26,27,28}, and to a lesser extent at other 
semiconductor-electrolyte 
interfaces: 
Cd$_3$P$_2$ (111)\cite{29}, CdIn$_2$S$_4$ (111)\cite{30}, GaN (001)\cite{31}, 
TiO$_2$\cite{32}. 
These 
studies revealed that the strength of the DC-electric field which could be applied 
electrochemically 
was limited by interface electrochemical reactions, such as oxidation of a silicon surface 
at anodic 
potential. To circumvent this restriction, EFISH generation studies were extended to 
Si-SiO$_2$ 
MOS structures with bias applied by a ring metal\cite{33} or semitransparent 
Cr\cite{19,34} 
gate 
electrode, and to GaAs-based MOS structures\cite{35}.
\par
A simple phenomenological model of EFISH based on the "interface field approximation" 
- which 
assumes linear dependence of the DC-field-induced nonlinear polarization on interface 
DCF 
strength and yields quadratic dependence of EFISH intensity on bias voltage - was 
developed for 
the Si-SiO$_2$-electrolyte interface in Refs. [25,26]. Since clear deviations from a 
quadratic bias 
dependence were observed\cite{33,34}, this model was improved by taking into account 
the 
nonlinear 
interference of DC-field induced and field-independent contributions to the nonlinear 
quadratic 
polarization as well as retardation and absorption effects\cite{34}. Further improvement 
resulted from 
considering the spatial inhomogeneity of the DCF and the DC-electric-field-induced 
contribution 
to the nonlinear polarization\cite{34}. These effects were later analyzed with a Green-function 
formalism\cite{36,37}. At present, the most comprehensive description of the EFISH 
phenomenon is 
presented in Ref. [37]. However, this analysis remains incomplete on three points. First, 
it is 
restricted to the depletion regime of the SCR, whereas experimentally applied biases 
have included 
accumulation and inversion regimes. Moreover, as we demonstrate in this paper, the 
transition 
from depletion to inversion to accumulation drastically changes the EFISH response. 
Second, 
surface quantization effects originating from strong field localization in inversion and 
accumulation 
regimes, as well as the role of interface states, should be taken into account. Third, 
multiple 
reflection interference in the SiO$_2$ layer, which significantly affects the SHG 
intensity from Si-SiO$_2$ structures\cite{38,39,40}, was neglected.
\par
In this paper we present a comprehensive phenomenological model of EFISH generation 
supported by experimental spectroscopic studies of p- and n-type Si(001)-SiO$_2$-Cr 
MOS 
structures. The key features of our model are: 1) a detailed electrophysical model of the 
SCR in 
the accumulation and inversion regimes, which takes into account interface states and 
oxide charge 
traps and their effect on the spatial DCF distribution in the SCR; 2) a rigorous nonlinear 
optical 
model of EFISH in the SCR, based on a Green-function formalism, which takes into 
account all 
retardation effects, absorption of the fundamental and SH radiation, multiple reflection 
interference of both the fundamental and SH waves in the oxide and optical interference 
between 
field-dependent and field-independent contributions to the quadratic nonlinear 
polarization. The 
key feature of our experiments is comprehensive observation of the dependence of SHG 
on 
numerous parameters, including applied bias, azimuthal sample rotation, wavelength near 
the 
direct two-photon $E_1$ transition, doping concentration, and oxide thickness. These 
combined 
dependences allow us to deconvolve the EFISH contribution fully from field-independent 
contributions. The non-quadratic bias dependence of the EFISH intensity predicted in 
Refs. 
[32,37], and its variation with doping concentration, oxide thickness, interfacial state 
density, and 
wavelength, is observed and analyzed in detail.

\section{ THEORETICAL BACKGROUND}
\subsection{Quadratic optical response of the Si-SiO$_2$ system}
\par
In the presence of a DCF the nonlinear polarization of a centrosymmetric semiconductor 
at the second-harmonic (SH) 
wavelength is given by\cite{41,42}:
\begin{equation}
{\bf P}^{NL}={\bf P}^S+{\bf P}^{BQ}+{\bf P}^{BD},
\label{eq1}
\end{equation}
where ${\bf P}^S$ is the surface nonlinear polarization, ${\bf P}^{BQ}$ is the bulk 
quadruple 
contribution, and ${\bf P}^{BD}$ is the bulk dipole DCF induced 
polarization. The 
last contribution is governed by the fourth-rank  cubic susceptibility tensor $\chi^{(3)}$ 
and can 
be written phenomenologically as
\begin{equation}
{\bf P}^{BD}=\chi ^{(3),BD}(2\omega ;\omega ,\omega ,0): %
{\bf E}(\omega ){\bf E}(\omega ){\bf E}_0,
\label{eq2}
\end{equation}
where ${\bf E}(\omega )$ and ${\bf E}_0$ are the amplitudes of the fundamental 
radiation and 
DCF strength inside the semiconductor, respectively. For crystals such as Si and Ge of 
point group 
symmetry $O_h$, $\chi ^{(3),BD}$ has 21 nonzero tensor components\cite{20}, of 
which 
only three 
are nonequivalent.
\par
The bulk quadrupole contribution in the plane-wave approximation is given by 
\begin{equation}
{\bf P}^{BQ}=\chi ^{(2),BQ}(2\omega ;\omega ,\omega): %
{\bf E}(\omega )i{\bf k}(\omega ){\bf E}(\omega ),
\label{eq3}
\end{equation}
where $\chi ^{(2),BQ}$ is a fourth-rank tensor which represents the quadrupole 
contribution to 
the quadratic nonlinear susceptibility from spatial dispersion and ${\bf k}(\omega )$ is 
the 
wavevector of the fundamental radiation in the semiconductor. $\chi ^{(2),BQ}$  has 
the same 
symmetry properties as $\chi ^{(3),BD}$. 
\par
For the surface contribution to ${\bf P}^{NL}$ the multipole expansion is hardly 
expected to be 
valid, and we suppose that\cite{43}:
\begin{equation}
{\bf P}^S=\chi ^{(2),S}(2\omega ;\omega ,\omega):{\bf E}(\omega ){\bf 
E}(\omega),
\label{eq4}
\end{equation}
where $\chi ^{(2),S}$  is a third-rank tensor representing an effective quadratic 
susceptibility of 
the surface layer, which includes a local part from breaking of inversion symmetry at the 
surface, 
and a nonlocal part from the discontinuity of the normal electric field component at the 
surface. 
The structure of $\chi ^{(2),S}$ depends on the particular crystalline face under 
consideration.
\par
The SH electromagnetic field ${\bf E}(2\omega )$ is found by solving the 
inhomogeneous wave 
equation for propagation of the SH wave with ${\bf P}^{NL}$ as a source 
term\cite{43,44}. 
The 
solution can be written formally in terms of the tensorial Green-function 
$\mathord{\buildrel{\lower3pt\hbox{$\scriptscriptstyle\leftrightarrow$}}\over G}(\bf {r},\bf {r}^\prime,{\rm 2} \omega)$, which is defined 
to be the solution of the wave equation with a point source at 
$\bf {r}^\prime$. 
Since translational symmetry in the interface plane is assumed, the SH field is given by
\begin{equation}
{\bf E}(z,k_{||},2\omega )=\int{\mathord{\buildrel{\lower3pt\hbox{$\scriptscriptstyle\leftrightarrow$}}\over G}
(z,z^\prime,k_{||},2\omega ){\bf P}^{NL}(z^\prime,2\omega )dz^\prime},
\label{eq5}
\end{equation}
where $k_{||}$ is the in-plane component of the SH wavevector. Hereafter we use the 
${\bf xyz}$ 
coordinate frame with the ${\bf xy}$  plane coinciding with the interface and the 
positive ${\bf 
z}$ -axis directed toward the semiconductor bulk. Expressions for the components of 
$\mathord{\buildrel{\lower3pt\hbox{$\scriptscriptstyle\leftrightarrow$}}\over G}$,  
are 
calculated in Refs. [43,44]. The DCF induced part of the SH field is given by:
\begin{eqnarray}
{\bf E}^{BD}(z,k_{||},2\omega )=F_{2\omega }F_\omega ^2\chi 
_{eff}^{BD}I_\omega {\bf p} \times \nonumber \\
\times \int\limits_0^{+\infty } {E_0(z^\prime)\exp \left( {i\left( {k_{2\omega ,z}+2k_{\omega 
,z}} \right)z^\prime} \right)dz^\prime}
\label{eq6}
\end{eqnarray}
where the scalar factor $\chi^{BD}_{eff}$ is a linear combination of components of 
$\chi^{(3),BD}$ which depends on the experimental geometry, $I_{\omega}$ is the 
intensity of 
the fundamental radiation, $k_{\omega,z}$ and $k_{2\omega,z}$ are the normal 
wavevector 
components of the fundamental and SHG radiation, respectively, in the semiconductor, 
the unit 
vector $\bf p$ defines the polarization of the EFISH field, and $F_{\omega}$ and 
$F_{2\omega}$ 
are the transmission factors which include Fresnel coefficients and a correction for 
multiple 
reflection  in the silicon oxide at both $\omega$ and $2\omega$. Eq.(\ref{eq6}) 
properly 
takes into 
account retardation, the penetration depth of the fundamental wave, the escape length of 
the SH 
wave and multiple reflection interference effects in oxide layer.

\subsection{DC-electric-field spatial distribution}
\par
To perform the integration in Eq.(\ref{eq6}) one must know the spatial distribution $E_0(z)$ 
across 
the SCR. In this section we consider the screening of this DCF within the framework of 
Fermi 
carrier statistics\cite{45,46,47}. The spatial distribution of the electrostatic potential 
$\varphi 
(z)$ in 
the planar semiconductor-dielectric system can be found as a solution of the 
one-dimensional 
Poisson equation:
\begin{equation}
{\partial  \over {\partial z}}\left( {\epsilon {\partial  \over {\partial z}}\varphi } 
\right)=4\pi n,
\label{eq7}
\end{equation}
where $\epsilon$ is the static dielectric constant of the semiconductor (dielectric) 
and $n=n(z)$ 
is the space charge density. The boundary conditions for Eq.(\ref{eq7}) are:
\begin{eqnarray}
\varphi (+\infty )=\mu, \nonumber \\
\varphi (-D)=\mu +\varphi _0,
\label{eq8}
\end{eqnarray}
where $\mu$ is the chemical potential of the semiconductor and $D$ is the thickness 
of the oxide 
film. The first equation in (\ref{eq8}) is a statement of charge neutrality in the bulk of 
the 
semiconductor. 
The second equation takes into account the application of external potential 
$\varphi_0$ to the 
metal electrode with respect to the semiconductor. We divide the charge density into 
field 
independent and field dependent terms:
\begin{equation}
n=n_{fi}+n_{fd},
\label{eq9}
\end{equation}
where $n_{fi}$ includes the density of the ionized donors $N_D$ and acceptors 
$N_A$, and fixed 
charge $n_{ox}$ trapped in the oxide layer near the semiconductor-dielectric interface:
\begin{equation}
n_{fi}=N_D+N_A+\delta (z+0)n_{ox},\,\,\,\,\,z\ge 0.
\label{eq10}
\end{equation}
Hereafter $z=+0$ and $z=-0$ denote positions near the interface just inside the 
semiconductor and just inside the dielectric, respectively. 
\par
The spatial distribution $n_{fd}(z)$ is, in principle, a nonlinear functional of the 
potential 
$\varphi$ at all points inside the semiconductor. However, first we find expressions for 
$n_{fd}(z)$  and $E_0(z)$ within the model of local screening of the DCF in a Fermi 
electron-hole gas, in which $n_{fd}(z)$  depends on the potential $\varphi$ at point $z$, 
$i.e.$. 
$n_{fd}(\varphi)=n_{fd}(\varphi(z))$. The field-dependent part of charge density 
consists of the 
concentration of holes $n_h$, electrons $n_e$, and interface traps $n_{it}$, which 
depend on the 
interface potential:
\begin{equation}
n_{fd}(z)=n_h(\varphi (z))+n_e(\varphi (z))+\delta (z-0)n_{it}(\varphi 
(z=+0)),\,\,\,\,\,z\ge 0.
\label{eq11}
\end{equation}
Since we assume that the SHG response comes from the semiconductor or 
semiconductor-
dielectric interface, we treat charges in the oxide layer as an effective fixed trapped 
charge 
$n_{ox}$. Since at $z>0$ the variable $z$ does not enter into Eq.(\ref{eq10},\ref{eq11}) 
explicitly and 
the charge 
density $n_{fd}$ depends on the coordinate via $\varphi(z)$, the Poisson equation 
(\ref{eq7}) 
has the first 
integral:
\begin{equation}
E_0^2(\varphi )= \frac{8\pi}{\epsilon}\int\limits_{\varphi+\mu} ^\mu {n(\varphi^\prime )}d\varphi^\prime.
\label{eq12}
\end{equation}
\par
Using the charge neutrality condition in the bulk of the semiconductor for completely 
ionized 
donors and acceptors yields
\begin{eqnarray}
N_D=eN_C\Phi \left ( \frac{\mu -\varepsilon _C}{kT} \right ),\\
N_A=-eN_V\Phi \left ( \frac{\varepsilon _V-\mu}{kT} \right ).
\label{eq13}
\end{eqnarray}

Eqs.(\ref{eq10},\ref{eq11}) have the form:
\begin{equation}
n_{fd}(\varphi (z))=eN_V\Phi \left( {{{\varepsilon _V-\varphi } \over {kT}}} \right)-
eN_C\Phi \left( {{{\varphi -\varepsilon _C} \over {kT}}} \right)+\delta (z+0)n_{it},
\label{eq14}
\end{equation}
\begin{equation}
n_{fi}=eN_C\Phi \left( {{{\mu -\varepsilon _C} \over {kT}}} \right)-eN_V\Phi 
\left( {{{\varepsilon _V-\mu } \over {kT}}} \right)+\delta (z-0)n_{ox},
\label{eq40}
\end{equation}
where
\begin{equation}
\Phi (\tau )={2 \over {\sqrt \pi }}\int\limits_0^\infty  {\sqrt x\left( {1+\exp 
\left( {x-\tau } \right)} 
\right)^{-1}dx},
\label{eq15}
\end{equation}
is the Fermi-Dirac integral; $N_V$ and $N_C$ are the density of states in valence and 
conduction bands, respectively, which depend on density-of-state and effective mass of 
electrons 
or holes; $\varepsilon _V$ and $\varepsilon _C$ are the energies of the upper level of 
the 
valence band and the lower level of the conduction band, respectively; $k$ is the 
Boltzmann 
constant, and $T$ is the temperature.
\par
Interface traps are charged mid-gap states at the semiconductor-dielectric interface 
resulting from 
interruption of the semiconductor lattice structure or interface imperfections. As the 
interface 
electrostatic potential changes, the trap levels move up or down while the Fermi level 
remains 
fixed. Interface trap density $n_{it}$ is defined in terms of the energy distribution 
$L^{A,D}(E)$ 
of trap levels across the semiconductor band gap:
\begin{eqnarray}
n_{it}(\varphi )=e\int\limits_{\varepsilon _V}^{\varepsilon _C} { ( L^D(E-
e\varphi )F^D(\mu -E+e\varphi )- L^A(E-e\varphi )F^A(\mu -E+e\varphi ) ) dE},
\label{eq16}
\end{eqnarray}
where superscripts $A,D$ denote acceptor or donor traps, and $F^A$ and $F^D$ are 
Fermi distribution functions:
\begin{equation}
F^{A,D}(\tau )=\left( {1+g^{A,D}\exp \left( {\mp {\tau  \over {kT}}} \right)} 
\right)^{-1},
\label{eq17}
\end{equation}
where coefficients $g^A=1/4$ and $g^D=2$ reflect the ground-state degeneracy of the 
acceptor and donor levels. The specific form of $L^{A,D}(E)$ depends on the preparation 
of 
the semiconductor-dielectric system. In the calculations we model this distribution as a 
set of 
Lorentz functions.
\par
Figure 1 shows the distributions $\varphi (z)$ and $E_0(z)$ across the SCR of p-doped 
silicon modelled within the above framework. In the depletion regime, where the 
Schottky 
approximation is valid, $\varphi (z)$ is close to a parabolic function. For larger applied 
bias 
corresponding to inversion, the SCR divides into a thin subsurface region of rapidly 
changing 
potential, and a long tail of gradually decreasing potential. The transition depth $z_0$ 
between 
these two regions lies at several nanometers. In the accumulation regime, $\varphi 
(z)$  drops 
completely within $z_0$.

\subsection{ The role of surface quantization effects in the subsurface region}
\par
The large gradient of the subsurface $\varphi (z)$   for accumulation and inversion 
regimes 
requires that quantum effects be considered in the screening of the DCF.  We take them 
into 
account via self-consistent calculations\cite{48}, using the Hartree-Fock (HF) approach 
to 
describe 
the exchange electron interaction. In the following we consider the screening of a 
"positive" (in the 
above notation) external potential in the subsurface region by electrons, with negligible 
contribution from holes.  The opposite case of "negative" potential is treated similarly. 
\par
The HF equation for the single-electron wave function $\psi _i(\bf r)$ is given by:
\begin{eqnarray}
\frac{\hat p^2}{2m}\psi _i({\bf r})+
e^2\sum\limits_{j\ne i} {<\psi_j({\bf r'})|(\epsilon |{\bf r}-{\bf r'}|)^{-1}
|\psi _j({\bf r'})>} \psi _i({\bf r})- \nonumber \\
-e^2\sum\limits_{j\ne i,||spins} {<\psi _j({\bf r'})|(\epsilon |{\bf r}-{\bf r'}|)^{-1}
|\psi _i({\bf r'})>}\psi _j({\bf r})+U_0({\bf r})\psi _i({\bf r})=E_i\psi _i({\bf r}),
\label{eq18}
\end{eqnarray}
where $U_0(z)=-E_0z+\int {n_{fi}({\bf r'})(\epsilon \,|{\bf r}-{\bf r'}|)^{-1}\,d^3r'}$ and the sum 
in the 
exchange (third) term is over states with parallel spins; brackets denote averaging over 
the 
stationary state. 
\par
Because of translation symmetry in the $x,y$-plane $\psi _i({\bf r})=\tilde \varphi 
_i(z)\,e^{i{\bf p}_i {\bf r}_{||}}$. We consider the case in which only one energy state for the 
subsurface 
electronic  -motion is responsible for most of the screening. This is confirmed by the 
numerical 
results. We also assume $\tilde \varphi _i(z)=\tilde \varphi (z)$ to be independent 
from $\bf 
p_i$. Then Eq.(\ref{eq18}) may be written in the form of a Schr{\"o}dinger equation 
with self-consistent 
potential $U_z$: 
\begin{equation}
E\tilde \varphi (z)=-\frac{\hbar ^2} {2m}\frac{\partial ^2}{\partial z^2}\tilde 
\varphi (z)-
U(z)\tilde \varphi (z),
\label{eq19}
\end{equation}
where
\begin{eqnarray}
U(z)=-E_0z+\int {\frac{n({\bf r'})C({\bf r},{\bf r'})+n_{fi}({\bf r'})}{\epsilon|{\bf r}-{\bf r'}|}d^3r'},\\
n({\bf r})=<\hat n({\bf r})>=e\sum\limits_K {\psi _K({\bf r})\psi _K^*({\bf r})}f({\rm E}_K),
\label{eq20}
\end{eqnarray}
$\hat n(\bf r)$ is the density operator, $f({\rm E}_k)=1/(1+e^{({\rm E}_k-\mu 
)/kT})$ is the 
Fermi occupation factor,
\begin{equation}
C({\bf r}-{\bf r'})=1-\frac{e^2} {n({\bf r})n({\bf r'})} \sum\limits_{j\ne i,||spins}
 {f({\rm E}_j)f({\rm E}_i)}e^{i({\bf p}_i-{\bf p}_j)({\bf r}-{\bf r'})}.
\label{eq21}
\end{equation}
$C({\bf r}-{\bf r'})$ can be interpreted physically as a correlation function for the in-plane motion 
of 
electrons. Boundary conditions for the wave function are given by
\begin{equation}
\tilde \varphi (0)=0,\,\,\,\,\tilde \varphi (z_0)=0.
\label{eq22}
\end{equation}
\par
From the equations above one can show that the potential $U(z)$ obeys the following 
equation 
for the 2D-system under consideration:
\begin{equation}
\frac{dU}{dz}=-E_0+\frac{2\pi } {\epsilon }\int {dz'(n(z')F(z-
z')+n_{fi}(z'){\rm sgn} (z-
z'))},
\label{eq23}
\end{equation}
where
\begin{equation}
F(z)=\int {\frac{{\rm sgn} (z)C(\rho |z|)} {\left( {1+\rho ^2} \right)^{3/2}}\rho d\rho }.
\label{eq36}
\end{equation}
The electrostatic potential $\varphi (z)$ obeys the equation
\begin{equation}
\frac{d\varphi }{dz}=\,-E_0+\frac{2\pi } { \varepsilon }\int 
{dz'(n(z')+n_{fi}(z'))}{\rm sgn} (z-z'),
\label{eq24}
\end{equation}
which can be derived from the Poisson equation. 
\par
Eqs.(\ref{eq23},\ref{eq24}) differ one from another by the factor $F(z)$ in Eq.(\ref{eq23}). 
It can be 
shown that 
since $C(r)\to 1$ at $r\to \infty $, $F(z)\to 1$ at $z\to \infty $. Therefore, the 
self-
consistent potential $U(z)$ is closely related to $\varphi (z)$. Moreover, remote 
charge 
layers contribute equally to $U(z)$  and $\varphi (z)$. Nevertheless, the electrostatic 
and 
self-consistent potentials are distinguished by the role of quantum-correlation effects in 
the 
electron plasma. The electrostatic potential describes interactions of a charged probe 
particle with 
other charges only via the electromagnetic field. The self-consistent potential $U(z)$   
also 
includes the electron's tendency to "wedge" itself into other electrons and repulse them 
via the 
exchange interaction, and therefore differs fundamentally from $\varphi (z)$. On the 
other hand, 
the EFISH bias dependence is expressed in terms of the classical potential $\varphi 
(z)$, 
because the major contribution to the semiconductor optical response comes from {\it 
bound} 
electrons, whereas screening in the semiconductor is caused by {\it free} carriers, and 
there are no 
correlation effects between these two different types of particles.
\par
Summarizing this section, we have obtained a set of Eqs.(\ref{eq19},\ref{eq20},\ref{eq21}) for the 
self-
consistent 
potential $U(z)$, electron wave function $\psi _i(\bf r)$ and charge density $n(z)$. 
These 
equations describe the screening in the immediate subsurface region $z<z_0$. This 
approach 
takes into account correlation effects in the electronic liquid via the factor $F(z)$ which 
is 
related to the in-plane correlation function $C({\bf r}_{||}-{\bf r}'_{||})$.

\subsection{ Numerical experiment }
\par
In this section the model bias dependence of the EFISH intensity is found by numerical 
integration 
of the Poisson equation and the wave equation. First, the dependence of the EFISH 
intensity on 
the doping of the semiconductor is taken into account. Then, the influence of parameters 
of the 
semiconductor-insulator interface on the amplitude of the EFISH wave is considered.
\par
To find the DCF induced SH field amplitude $E^{BD}$ for applied bias $U$, $E_0(z)$ 
has 
been calculated by numerically solving the first integral of the Poisson Equation 
(\ref{eq7}) with 
the 
charge densities given by Eqs.(\ref{eq14},\ref{eq40}). The boundary condition at the surface 
of 
the metal 
electrode of MOS structure is given by $\varphi _0=U$. $U$ is related to the 
interface field 
$E_{int}=E_0(z=+0)$ and interface potential $\varphi _{int }=\varphi (z=+0)$, 
by
\begin{equation}
U=\epsilon _{sc}\epsilon_d^{-1}E_{int}(\varphi _{int })D+\varphi _{int}. 
\label{eq25}
\end{equation}
The parameters of silicon, which is used as a model semiconductor, have been taken 
from Ref. 
[49]. According to Eq.(\ref{eq6}), the EFISH field $E^{BD}$ is a product of the integral 
\begin{equation}
I(U)\equiv I_1+iI_2=\int\limits_0^{+\infty } {E_0(z)\exp \left( {i\left( {k_{2\omega 
,z}+2k_{\omega ,z}} \right)z} \right)dz}, 
\label{eq38}
\end{equation}
and the complex factor $F_{2\omega }F_\omega ^2\chi 
_{eff}^{BD}$, 
which is a bias-independent constant for a given fundamental wavelength. This allows us 
to 
neglect the complex value of the latter term and simulate the bias dependence of 
$I_1(U)$ and 
$I_2(U)$ by the bias dependence of ${\rm Re} E^{BD}$ and ${\rm Im} E^{BD}$ in units of 
$\left| 
{F_{2\omega }F_\omega ^2\chi _{eff}^{BD}} \right|$. This notation is used in the 
numerical 
experiment shown in Figs. 2,4 and 5. Figure 2 shows ${\rm Re} E^{BD}$ and ${\rm Im} 
E^{BD}$ as 
functions of the bias applied to the MOS structure, calculated by evaluating the integral 
in Eq.(\ref{eq6}) 
for different dopant concentrations of a n-type silicon wafer covered by silicon oxide film 
19 nm 
thick. The fundamental radiation wavelength is presumed to be 730 nm. 
\par	
Two important trends in these curves are noteworthy. First, ${\rm Im} E^{BD}$  
depends 
strongly on the bias only in the region of negative biases between 0 V and a saturation 
bias we 
denote as $U_0$. Outside of this interval the amplitude of the EFISH field saturates. 
This strongly 
contradicts the previous phenomenological assumption that the amplitude of the EFISH 
field 
depends linearly (and the EFISH intensity quadratically) on the applied bias. The 
saturation of the imaginary part of the 
EFISH field amplitude for $U<U_0$ and $U>0$ is attributed to the inversion and 
accumulation regimes of the external bias screening in the SCR (see inset in Fig. 1) as 
the DCF is 
mostly localized inside a thin subsurface layer of nm-scale thickness. Since the imaginary 
part of 
the Green's function is equal to zero exactly at the interface, ${\rm Im} E^{BD}$  becomes 
practically insensitive to the DCF inside the inversion and accumulation layers. Thus 
$U=U_0$ 
and $U=0$ define end-points of a bias region which corresponds to the depletion regime; 
the 
interface potential $\varphi _0$ for external bias $U_0$ is equal to $2(\varepsilon _i-
\mu 
)$, where $\varepsilon _i$ is the midgap energy\cite{50}.
\par
Second, decrease of dopant concentration leads to the decrease of the absolute value of 
$U_0$  
and $E^{BD}$. Figure 3 shows the dependence of the absolute value of $U_0$  on the 
dopant concentration of the n-type silicon wafer for various oxide thicknesses. For dopant 
concentrations larger than $10^{16}\,{\rm cm}^{-3}$, the absolute value of $U_0$  scales 
approximately as the square root of $N_D$, while for smaller doping levels $\left| 
{U_0} 
\right|$ scales as $\ln N_D$, as is clearly shown in the inset of Fig. 3. Applied bias 
$U_0$  
according to Eq.(\ref{eq25}) consists of voltage drops $\epsilon _{sc}\epsilon _d^{-
1}E_0(\varphi _0)D$ across the oxide film, and $\varphi _0$ across the silicon SCR. 
Within 
the Schottky approximation for the SCR\cite{47}, the interface potential $\varphi _0$  
and 
interface 
field $E_0$, corresponding to applied bias $U_0$   are given by
\begin{equation}
E_0=2\sqrt {\xi \varphi _0},\,\,\varphi _0=2(\varepsilon _i-kT\ln(N_DN_C^{-
1})),\,\,\xi =2\pi eN_D\epsilon_{sc}^{-1}.
\label{eq26}
\end{equation}
Therefore, for high doping levels the applied bias mostly drops across the oxide layer and 
$U_0(N_D)\propto E_0(N_D)\propto \sqrt {N_D\ln (N_DN_C^{-1})}$. For low 
doping the 
interface potential dominates and $U_0(N_D)\propto \varphi _0(N_D)\propto \ln 
(N_DN_C^{-
1})$. For thiner oxide layers, less of the applied voltage is dropped across the oxide 
and the 
transition from a logarithmic to a square root doping dependence of $U_0$ occurs at a 
higher 
doping level.
\par
Figure 4 shows the EFISH amplitudes for applied bias $U_0$  as functions of donor 
concentration, $N_D$. Over a wide range of concentrations ${\rm Re} E^{BD}$ and 
${\rm Im} E^{BD}$ depend on the square root of $N_D$. The latter can be explained by 
integrating Eq.(\ref{eq6}) with a linear DCF $E(z)=E_0-2\xi z$ across the SCR, as in the Schottky model. 
This 
integration yields the following expressions for the EFISH field:
\begin{equation}
{\rm Re} E^{BD}\propto E_0\Delta _2+2\xi \frac{\Delta _1^2-\Delta _2^2}{\Delta 
_1^2+\Delta_2^2},
\label{eq27}
\end{equation}
\begin{equation}
{\rm Im} E^{BD}\propto E_0\Delta _1-4\xi \frac{\Delta _1\Delta _2}{\Delta 
_1^2+\Delta_2^2},
\label{eq28}
\end{equation}
where $\Delta _1={\rm Re} (2k_\omega +k_{2\omega })$ and $\Delta _2={\rm Im} 
(2k_\omega 
+k_{2\omega })$. Since the interface field $E_0=2\xi W$ depends linearly on the 
width of the 
SCR, the restrictions $W\Delta _1>>1$ and $W\Delta _2>>1$ lead to the following 
expression for the complex SH field: $E^{BD}\propto E_0(\Delta _2+i\Delta 
_1)\propto \sqrt 
{N_D\ln (N_DN_C^{-1})}$.  Thus ${\rm Re} E^{BD}$ and ${\rm Im} E^{BD}$ scale 
approximately 
as the square root of $N_D$.  Furthermore, ${\rm Im} E^{BD}/{\rm Re} E^{BD}=\Delta 
_1\Delta _2^{-
1}$, i.e. the ratio of  ${\rm Re} E^{BD}$  to ${\rm Im} E^{BD}$  is the ratio of the 
characteristic 
length scale of absorption, $\Delta^{-1}_2$, to that of retardation, $\Delta^{-1}_1$, for 
the SH 
waves. As the energy of the 365 nm SH photon used in our calculations is close to the 
$E_1$ 
critical point of silicon, $\Delta_1=(1/5.7)\, {\rm nm}^{-1}$ and $\Delta_2=(1/21.5)\, {\rm nm}^{-1}$ 
are 
sufficiently 
large to satisfy the conditions $W\Delta _1>>1$ and $W\Delta _2>>1$ for dopant 
concentrations up to $10^{18}\, {\rm cm}^{-3}$ (Fig. 4). As shown in the inset of Fig. 4, the 
ratio ${\rm Im} 
E^{BD}/{\rm Re} E^{BD}$ is close to the value of $\Delta _1\Delta _2^{-1}=3.89$.

\subsection{The role of interface states in the EFISH phenomenon}
\par
A sheet of charged interface states changes the relationship between a potential drop 
across silicon 
and an applied bias due to the boundary condition for normal components of the electric 
displacement vector $\bf D$. To demonstrate the role of interface traps in the EFISH 
phenomenon 
we consider the distribution of trap levels across the silicon band gap as a set of 
Lorentz's 
functions. The charge density of interface traps $n_{it}$ as a function of the interface 
electrostatic potential is given by 
\begin{eqnarray}
n_{it}(\varphi (z=0))=e\int\limits_{\varepsilon _V}^{\varepsilon _C} 
{dE\sum\limits_M {{\rm sgn} \left( n_{it}^M \right ) }F^M(\mu +e\varphi -E)}\times \nonumber \\ 
  \times \sum\limits_j {N_{M,j}\delta _{M,j}^2\left( {\delta _{M,j}^2+\left( 
{E-\varepsilon _{0M,j}} \right)^2} \right)^{-1}}, 
\label{eq29}
\end{eqnarray}
where $M=A,D$, $j$ numerates Lorentz functions of the energy distribution of the 
trap levels. 
$N_{M,j},\delta _{M,j}$ and $\varepsilon _{0M,j}$ denote the effective number of 
traps per 
unit area, the width and central position of $j$-th Lorentz peak, respectively. These 
Lorentz 
functions simulate the continuous energy distribution of traps. By setting $\delta 
_{M,j}\to 0$ 
one can account for discrete levels.
\par
Figure 5 shows the bias dependence of ${\rm Re} E^{BD}$ and ${\rm Im} E^{BD}$ for the MOS 
structure used in our experiment which is comprised of  n-type Si with a donor 
concentration of 
$10^{18}\, {\rm cm}^{-3}$ and a 19-nm thick thermal SiO$_2$ layer. Interface traps are 
presumed to be 
acceptors with $N_A=10^{13}\,traps\cdot cm^{-2}eV^{-1}$ and $\delta 
_A=0.5\,\,kT$. 
Different central positions are considered: $\varepsilon _{0A}=\mu $ (thin lines), 
$\varepsilon 
_{0A}-\mu =-10\,\,kT$ (thick lines) and 
$\varepsilon _{0A}-\mu =-20\,kT$ (dashed lines). The distribution of such traps 
across the 
silicon band gap is sketched at the inset in Figure 5. The dotted lines are presented for 
comparison 
to the same field components in the absence of traps. For negative biases in the 
inversion regime, 
bands are bent in such a way that all the trap levels are above the Fermi energy, 
acceptor traps are 
empty and bias dependence of the SH field components is unaffected by the presence of 
these 
uncharged traps.  As the magnitude of the negative bias is decreased, the bands are bent 
less and 
trap levels begin to fall below the Fermi energy, first the traps with low energy levels, 
then those 
with higher energy. Consequently, the bias dependence of ${\rm Re} E^{BD}$ and ${\rm Im} 
E^{BD}$ 
for low energy acceptor traps starts to deviate from the dependence for $N_A=0$, 
demonstrating the saturation-like feature. This is attributed to the pinning of the Fermi 
level. As 
the level of the neutral traps crosses the Fermi energy, the charge density of interface 
traps   
changes and application of a smaller bias leads to a decrease in the voltage drop across 
the oxide 
film while the interface potential and the DCF spatial distribution remain fixed until the 
trap level is 
completely filled. The bias dependence of ${\rm Re} E^{BD}$ and ${\rm Im} E^{BD}$ for 
$N_A\ne 0$ passes through the SHG intensity zero-point for a flatband voltage 
$U_{fb}$, which 
depends linearly on the interface charge $n_{it}$. In the case of donor interface traps 
the same 
effects are obtained, but the bias dependence of ${\rm Re} E^{BD}$ and ${\rm Im} E^{BD}$ for 
$N_D=0$ and $N_D\ne 0$ differ in the inversion regime.

\section{Results and Discussion}
\subsection{Experimental}
\par
For the EFISH experiments the output of an unamplified Ti-Sapphire laser ranging from 
710 to 800 nm was used. The Ti-sapphire laser generates 120-fs pulses with average 
power of 200 to 300 mW, which is well below the damage threshold of the 
semiconductor. The p-polarized beam was focused onto the sample at a $45^0$ angle of 
incidence. Reflected p-polarized SHG signal was selected by the use of appropriate filters 
and directed into a photon-counting system.  High intensity, high repetition rate, short 
pulses provided a good signal-to-noise ratio in our experiments while avoiding significant 
sample heating. A small split off portion of the fundamental beam was focussed through 
a $z$-cut quartz crystal that provided a reference SHG signal.
\par
The MOS structures were fabricated from two types of Si(001) wafers: (I) - a highly-
doped n-type ($10^{18}\, {\rm cm}^{-3}$, Sb doped ) wafer covered by a 19 nm thick SiO$_2$ 
film, 
and (II) - a low-doped p-type ($1.5 \cdot 10^{15}\, {\rm cm}^{-3}$, B doped) wafer with a 8.7 
nm thick SiO$_2$ film. A 3 nm semitransparent chromium cap layer, and an ohmic 
aluminum backside electrode were evaporated onto the samples. Single-wavelength 
ellipsometry was used to measure the SiO$_2$ thicknesses. As an independent 
calibration of the flatband voltage, spatially resolved surface photovoltage measurements 
were performed on the same samples. The external bias voltage was applied between the 
chromium and aluminum electrodes. The SHG response from the chromium layer was 
verified to be negligible in comparison with the SHG signal from the buried Si(001)-
SiO$_2$ interface. 
\par
The bias dependence of the rotational azimuthal anisotropy of the EFISH intensity was 
measured over a wide range of the bias voltages at various fundamental wavelengths 
from 710 nm to 800 nm. Figure 6 shows the azimuthal dependence of the EFISH 
intensity measured for an n-Si(001) MOS structure. The pronounced four-fold symmetric 
anisotropy of the EFISH intensity superimposed on a significant isotropic (i.e. 
independent from the azimuthal angle) background was observed at most biases. 
Variation of the applied voltage changes the amplitudes of both the four-fold symmetric 
and isotropic contributions, both of which increase with increasing the absolute value of 
the bias. At the center of the applied bias region near -2.75 V (upper panel) azimuthal 
dependence possesses a significant eight-fold symmetric component, which appears to be 
comparable with isotropic and four-fold components for the same bias. As the applied 
voltage passes through this bias the phase of the anisotropy shifts by $\pi/4$. Similar 
features of the field-induced rotational anisotropy were observed throughout the studied 
spectral range. Figure 7 shows the azimuthal dependence for a p-MOS structure which 
demonstrates similar behavior, except that the eight-fold symmetric component is 
observed at -1.2 V and the isotropic component is appeared to be quite larger than the 
four-fold one.

\subsection{EFISH at Si(001)-SiO$_2$ interface: Role of the spatial DCF distribution}
\par
The azimuthal angular dependence of the SHG intensity from the Si(001)-
SiO$_2$ interface in the presence of the DCF can be described phenomenologically as 
optical interference of DC-field dependent, isotropic and DC-field independent, four-fold 
symmetric components of the SH field:
\begin{eqnarray}
I_{2\omega}(\psi ,V)=\left| {a(V)+b\cos (4(\psi -\psi _0))} \right|^2= \nonumber \\
=c_0\left(V\right)+c_4\left(V\right)\cos(4(\psi-\psi_0))+c_8\cos(8(\psi-\psi_0)),
\label{eq30}
\end{eqnarray}
where $\psi_0$ is the azimuthal angle of a maximum of rotational anisotropy, $a$ and 
$b$ are the amplitudes of isotropic and anisotropic components of the SH field. The 
surface, ${\bf P}^S$, and the bulk DCF induced, ${\bf P}^{BD}$, components of the 
nonlinear polarization, ${\bf P}^{NL}$, contribute to the isotropic component $a$ while 
the four-fold symmetric component originates from the bulk quadruple polarization, 
${\bf P}^{BQ}$. For the sake of simplicity we put the amplitude $b$ of the four-fold 
symmetric anisotropic component as a real quantity and define the phase of the isotropic 
component $a=a^\prime+ia^{\prime\prime}$ with respect to $b$. As a result the dependence of the EFISH 
intensity on the azimuthal angle, $\psi$, is given by a Fourier expansion (\ref{eq30}) 
with 0-th, 4-th and 8-th Fourier components:
\begin{equation}
c_0=a^{\prime 2}+a^{\prime\prime 2}+\frac{1}{2}b^2,\,\, c_4=2a^\prime b,\,\, c_8=\frac{1}{2}b^2.
\label{eq31}
\end{equation}
Figure 8 shows the bias dependence of the isotropic Fourier component of the EFISH 
intensity  (left panel) and of the normalized four-fold Fourier component 
$c_4 \left(2\sqrt {2c_8} \right) ^{-1}$ (right panel), which is exactly the $a'$ 
component of the 
EFISH field. The eight-fold symmetric component, $c_8$, appears to be field-
independent throughout the range of applied biases. The error bars are the averaged 
amplitudes of Fourier components $c_1$ and $c_3 \left( 2\sqrt {2c_8} \right) ^{-1}$. 
The 
component $c_0(U)$ is quadratic with a minimum at -3.1 V. The $a'$ component 
passes through zero-point also at about -3.1 V and depends on bias nearly linearly with 
pronounced deviations from linearity at the edges of the bias range. These bias 
dependences have been fitted within the model described above with the amplitude of the 
field-independent part of $a$ and the flatband voltage $U_{fb}$ as adjustable 
parameters. Figure 8 shows the model results with $U_{fb}=0.7\, {\rm V}$ by solid curves 
which agree well with the experimental data. The obtained value of the flatband voltage 
significantly differs from either minimum of $c_0(U)$ or the bias for which $a'=0$. 
This difference is attributed to the optical interference of the DCF dependent (bulk) and 
DCF independent (surface) contributions to $a$. For this highly doped MOS structure 
the entire 8 V range of applied biases of corresponds to the depletion regime. 
\par
Figures 9 and 10 show the bias dependence of $c_0$ and $a'$ for the p-MOS structure.  
The quadratic behavior of $c_0(U)$ with a minimum at -1.25 V and approximately 
linear dependence of $a'(U)$ with deviations at the limits are similar to the trends of 
the n-MOS structure. The model of the experimental data with dopant concentration of 
$1.5 \cdot 10^{15}\, {\rm cm}^{-3}$, shows a clear step-like feature near the center bias which 
corresponds to the depletion regime of the SCR in p-type silicon. However, such 
peculiarity has not been observed experimentally. This discrepancy between the model 
and experiment occurs for the surface-quantization calculation as well, because at these 
small biases the surface-quantization effects are not of importance. 
\par
One possible explanation for the experimentally measured bias dependences is the 
influence of photoinduced effects on the EFISH intensity. The absorption of femtosecond 
laser pulses leads to the excitation of electron - hole pairs in the SCR. The DCF in the 
SCR separates these photo-induced carriers and the density of the charge injected into 
SCR for the pulse duration $\tau \sim 120\,{\rm fs}$ is on the order of $10^{17}\, {\rm cm}^{-
3}$\cite{6}. 
The presence of these extra charges should lead to a decrease of the SCR width. 
Systematic theoretical description of this effect requires a model that rigorously accounts 
for the kinetics of electron-hole recombination in the subsurface layer.  In our model the 
parameter responsible for the SCR width is the dopant concentration. Therefore, 
photoinduced effects can be {\it effectively} taken into account by variation of $N_D$. 
Thick curves in Figures 9 and 10 show the fit with $N_D=10^{17}\, {\rm cm}^{-3}$, which is 
two orders of magnitude larger than the actual dopant concentration. For such a doping 
level the range of biases which correspond to the depletion regime is sufficiently larger 
(about 4 V) and the transition from depletion to inversion and accumulation occurs more 
gradually than for a lower doping level. Much better agreement of the model with 
experimental data is achieved.
\par
For biases larger then 4 V, which correspond to the strong inversion regime, clear 
deviations of the model from experimental data are obtained. This is attributed to the 
strong localization of DCF inside a very thin subsurface layer where the bulk description 
of the DCF screening is hardly expected to be valid and one should take into account 
surface quantization effects. The dashed curve shows the approximation of the data by 
the model with quantum corrections, which demonstrates a better agreement with 
experimental data points in this bias region.

\subsection{The EFISH spectroscopy: Bulk origin of DC-field-induced contribution}
\par
Tuning the fundamental wavelength in the vicinity of the direct two-photon $E_1$ 
transition allows measurement of the spectrum of the EFISH intensity and deconvolution 
of the bulk and red-shifted surface contributions to the SHG signal\cite{9,11,51}. Figure 
11 
shows the bias dependence $c_4(U)$ for various wavelengths of the fundamental 
radiation, $\lambda_{\omega}$. Tuning of $\lambda_{\omega}$ from 800 nm to the 
two-photon resonance  near 3.4 eV ($\lambda_{\omega}=730\, {\rm nm}$) produces stronger 
bias dependence of both $c_4(U)$ and $c_0(U)$. Further decrease of 
$\lambda_{\omega}$ results in a reduced bias dependence. The bulk quadrupole 
component of the SH field, $b \equiv E_{anis}^{BQ}$, contributes to both the isotropic, 
$c_0$, and the four-fold symmetric, $c_4$, Fourier components. To extract the spectral 
dependence of the EFISH field, $E^{BD}$, one must find the spectrum of $\left| 
E^{BQ}_{anis}\right |$. The latter has been obtained from the spectrum of the eighth 
Fourier component $c_8$ averaged over the entire bias region. Integration of the 
product of the Green's function and the bulk quadrupole polarization, according to Eq. 
(\ref{eq5}), gives the spectral behavior of $\left| E^{BQ}_{anis}\right |$ in the 
form: 
\begin{eqnarray}
\left| E_{anis}^{BQ}(\Omega ) \right|=I_\omega \left| \frac{k_{\omega 
,z}^2(\Omega )}{k_{2\omega ,z}(\Omega )+2k_{\omega ,z}(\Omega )} \right|\left| 
F_\omega ^2(\Omega )F_{2\omega }(\Omega ) \right| \left| \chi^{(2),BQ}(\Omega ) \right|.
\label{eq32}
\end{eqnarray}
Hereafter, $\left| {\chi ^{(2),BQ}(\Omega )} \right|$ is the magnitude of a 
combination of $\chi ^{(2),BQ}$ tensor components responsible for the four-fold 
symmetric part of ${\bf P}^{BQ}$. $I_\omega$ is the fundamental intensity. Figure 
12 shows the spectrum of the magnitude of the effective quadruple susceptibility $\left| 
{\chi ^{(2),BQ}(\Omega )} \right|$. The filled symbols in Figure 12 show the spectral 
dependence of the effective cubic susceptibility $\chi ^{(3),BQ}$ extracted from the set 
of the bias dependences $c_0(V,\lambda)$. Both spectral dependences of $\left| {\chi 
^{(2),BQ}(\Omega )} \right|$ and $\left| {\chi ^{(3),BD}(\Omega )} \right|$  peak 
at approximately 3.4 eV and have been fitted by a single Lorentz function with a real 
spectral background:
\begin{equation}
\chi ^M(\Omega )=\alpha +\frac{\beta}{\Omega-\omega_M+i\delta},
\label{eq33}
\end{equation}
with $M=BQ,BD$. The solid curves in Figure 12 show the spectral fits of $\left| 
{\chi ^{(2),BQ}(\Omega )} \right|$  and $\left| {\chi ^{(3),BD}(\Omega )} 
\right|$   by Eq.(\ref{eq33}) with the parameters presented at Table I. The values of 
resonance positions obtained are shown to be close to 3.38 eV. This is consistent with 
the energy of the bulk $E_1$ critical point as known from linear spectroscopy and fully 
indicates a bulk origin of the EFISH response. 

\subsection{Low-frequency electromodulation SHG spectroscopy of Si(001)-SiO$_2$ interface}
\par
Modulation techniques are widely used in optical spectroscopy\cite{52} because of their 
sensitivity. The right side of Fig. 13 shows the schematic of the low-frequency electromodulation 
of the SHG signal from Si-SiO$_2$ interface in a MOS structure by the application of the 
superposition of DC-bias $U$ and low-frequency squarewave modulation voltage $\Delta 
U(\Omega)$ with the amplitude $\Delta U$ and frequency $\Omega$. Microwave 
frequency and pulse-voltage modulation of the SHG response in Si-based MOS structures were 
studied in Refs.[37] and [33], respectively. Low-frequency electromodulation SHG from GaN 
surface in electrochemical cell was studied in Ref. [31].
\par
The efficiency of the modulated SHG signal $\alpha (U, \Omega )$ at certain DC-bias $U$ can be 
defined by a relative increment of the EFISH intensity while applying the modulation voltage 
$\Delta U\left( \Omega  \right)$:
\begin{eqnarray}
\alpha (U,\Omega )=\frac{2\left( I_{2\omega }(U+\Delta U)-I_{2\omega }(U-\Delta U) 
\right)}{\left( I_{2\omega }(U+\Delta U)+I_{2\omega }(U-\Delta U) \right)} \approx 
\frac{dI_{2\omega }(U)}{dU}\frac{\Delta U\left( \Omega  \right)}{I_{2\omega }},
\label{eq34}
\end{eqnarray}
and appears to be a differential characteristic of the EFISH phenomenon which is complementary 
to the static EFISH dependence $I_{2\omega }\left( U \right)$.
\par
Figure 13 shows the experimental {\it static} (DCF-induced) EFISH bias dependence 
$I_{2\omega }\left( {U+\Delta U} \right)$ measured at p-Si(001) MOS structure for $\Delta 
U=0.6$ V which is typically featureless for the DCF induced SHG. Figure 14 shows the 
experimental bias dependence $\alpha (U)$ (open symbols) of the efficiency of the modulated 
EFISH for for $\Delta U=0.6$ V, and $\Omega =100$ Hz at an azimuthal angle $\psi=0$ that 
minimizes the anisotropic EFISH intensity. This dependence shows a flat feature in the vicinity of 
$U=0$ V. This feature is not seen on the bias dependence of the numerical derivative of 
$I_{2\omega }\left( U \right)$ function shown in Fig. 14 (solid symbols). The bias dependence 
$\alpha (U)$ as calculated in the framework of the phenomenological model above (Sections A 
and B in Part II) is presented in Fig. 14 (solid curve). This curve does not show a flat feature in the 
vicinity of $U=0$. This flat-like feature in the experimental differential bias dependence is likely 
coming from the large modulation amplitude, which could not be reduced for technical reasons.
\par
The inset in Fig. 12 shows the spectral dependence of the modulation efficiency in the tuning 
region of the Ti:Sapphire laser. A peak in the spectral dependence $\alpha (\lambda _\omega )$ is 
observed at the two-photon energy $2\hbar \omega = 3.41$ eV with half-width $\hbar \Delta 
\omega = 0.023$ eV. The spectral position of this peak is close to the bulk $E_1$ resonance. This 
confirms once again the bulk origin of the DCF-induced term of the nonlinear polarization in Eqs. 
(1) and (2). The spectral half-width of the resonance in the differential response $\alpha (\lambda 
_\omega )$ is smaller then the half-width of the resonances of the electrostatic (DCF-induced) 
EFISH terms in Fig. 12. This shows the increased sensitivity of the EFISH modulation spectra to 
density of states in the semiconductor valence and conduction bands.

\section{Conclusions}
\par
In summary, the DC-electric-field-induced SHG and the low-frequency electromodulation SHG 
spectroscopy of Si(001)-SiO$_2$ interfaces in p- and n-type Si(001)-SiO$_2$-Cr MOS structures 
have been studied. The dependences of the DC-electric-field-induced SHG intensity on the applied 
bias are shown to be sensitive to the doping concentration of silicon, oxide thickness and 
fundamental and SHG wavelengths. From spectroscopy of the anisotropic EFISH dependences the 
field-induced contribution has been extracted and the spectrum of the cubic susceptibility 
$\chi^{(3)}$ appears to be peaked at the energy of the bulk $E_1$ critical point. The presence of 
significant EFISH contribution at an unbiased Si(001)-SiO$_2$ interface due to the initial band 
bending has been observed. This initial band bending contribution should be taken into account in 
the further interpretation of the spectroscopic SHG measurements at Si(001)-SiO$_2$ 
interfaces\cite{9,51}.
\par
A general phenomenological model of the EFISH phenomenon is developed. This includes a 
comprehensive analyses of the generation and the propagation of the EFISH wave in the silicon 
space charge region taking into consideration the retardation and absorption effects, optical 
interference of the DC-field-dependent and DC-field-independent contributions to the SH waves 
and interference of multiple reflections in the oxide layer. The spatial distribution of the DC-field-
induced bulk dipole nonlinear polarization is calculated using the rigorous DCF distribution across 
the SCR taking into account surface quantization effects. The influence of the silicon doping level, 
oxide thickness, interface states and oxide charge traps on the screening of the external DCF in the 
SCR is studied. We have demonstrated the sensitivity of the EFISH probe to the charge 
characteristics of the Si(001)-SiO$_2$ interface which makes this technique promising as a 
noninvasive sensor of the MOS devices for the mapping of interface charge distribution.

\section{ACKNOWLEDGMENTS}
\par
We are grateful to L. V. Keldysh, J. K. Lowell, H. van Driel, T. F Heinz, G. L{\"u}pke, U. 
H{\"o}fer, W. 
Daum, J. F. McGilp, A. Yodh, H. W. K. Tom, D. von der Linde, Th. Rasing, O. Keller, and K. 
Pedersen for valuable discussions.
\par
This work was supported by the U.S. AFOSR (Contract F49620-95-C-0045), the Robert Welch 
Foundation (Grant F-1038), the Texas Advanced Technology Program (ATPD-354), and the U.S. 
National Science Foundation Science and Technology Center Program (Grant CHE-8920120), 
INTAS-93 grant 0370(ext), special Grant 96-15-96420 from the Russian Foundation of Basic 
Research (RFBR) for the Leading Russian Scientific Schools, RFBR grants 97-02-17919, 97-02-
17923, Russian Federal Integration Program "Center of Fundamental Optics and 
Spectroscopy" and Programs of Russian Ministry of Science "Physics of Solid State 
Nanostructures" and "Fundamental Metrology".

\newpage
{\bf Figure and Table Captions}
\vspace{2mm}
\begin{tabular}{ccccc}
\hline
&$\alpha$, rel.un. & $\beta$,rel.un. & $\delta$,eV & $\omega _M$,eV \\
\hline 
$BQ$ & -0.005 & 0.006 & 0.054 & 3.382\\

$BD$ & -1.216 & 0.378 & 0.053 & 3.384\\
\hline
\end{tabular}

\noindent
Spectral parameters of $\left| {\chi ^{(2),BQ}(\Omega )} \right|$ and $\left| \chi 
^{(3),BD}(\Omega ) \right|$.
\vspace{2mm}

\noindent Fig. 1:
The spatial electrostatic potential (left panel) and DC-electric field (right panel) 
distribution across 
the SCR of p-doped silicon (doping concentration of $1.5 \cdot 10^{15}\, {\rm cm}^{-3}$) for 
different 
values of interface potential: +0.95 V (inversion), +0.6 V (depletion) and -0.33 V 
(accumulation). 
The upper panel is the sketch of the potential and field distribution across the MOS 
structure.

\noindent Fig. 2:
The bias dependences of the real (left panel) and imaginary (right panel) parts of the 
EFISH field 
$E^{BD}$ for different doping levels of silicon wafer. Parameters of MOS structure used 
are in 
the text. The flatband 
voltage is 
supposed to be zero. Insets: the bias dependences in the vicinity of zero-point bias. 

\noindent Fig. 3:
The absolute value of the depletion bias $U_0$ vs. the doping concentration $N_D$ of the n-Si wafer for Si-
SiO$_2$-
metal MOS structures with different SiO$_2$ thicknesses: 1 nm (filled squares), 8.7 nm 
(open 
squares), 19 nm (filled circles), and 50 nm (open circles). Solid curves are guides to the 
eye. Inset: 
dependence $U_0(N_D)$ for MOS structure with 1 nm thick oxide in the linear scale.

\noindent Fig. 4:
The real (open circles) and imaginary (filled circles) part of $E^{BD}$ for the depletion 
bias 
$U_0$ vs. the doping concentration $N_D$ of the silicon wafer with 19 nm thick oxide film.
 Solid curves are guides to the 
eye. Inset: the doping dependence of the ratio of ${\rm Im} ( E^{BD} ) /{\rm Re} ( E^{BD}) $.

\noindent Fig. 5:
The bias dependences of the real (left panel) and imaginary (right panel) part of the 
EFISH field 
$E^{BD}$ for different parameters of acceptor interface states. The energy spectrum of 
interface 
states is simulated by Lorenzian function with density of $N_A=10^{13}\, {\rm traps \cdot cm}^{-2} 
{\rm eV}^{-1}$, width of $\delta _A=0.5\,\,kT$, and different central positions - 
$\varepsilon 
_{0A}=\mu $ (thin curves), $\varepsilon _{0A}-\mu =-10\,\,kT$ (thick curves) and 
$\varepsilon 
_{0A}-\mu =-20\,kT$ (dashed curves) and sketches on the inset. The dotted curves 
present for 
comparison the same dependences without traps. Parameters of the MOS structure are 
in the text.

\noindent Fig. 6:
p-in,p-out SHG signal from n-Si(001) MOS structure at several biases for $\lambda 
_\omega=725\, {\rm nm}$ 
($2\hbar \omega =3.43\, {\rm eV}$) vs. sample azimuthal angle. Solid curves are fits to 
data by the 
0-th, 4-th and 8-th Fourier components. 

\noindent Fig. 7:
p-in,p-out SHG signal from p-Si(001) MOS structure at several biases for for $\lambda 
_\omega=730\, {\rm nm}$ ($2\hbar \omega =3.41\,\,{\rm eV}$) as a function of sample azimuthal 
angle. Solid 
curves are fits to data by the 0-th, 4-th and 8-th Fourier components. 

\noindent Fig. 8:
Bias dependences of isotropic $c_0$ and normalized four-fold $a'=c_4\left( {2\sqrt 
{2c_8}} 
\right)^{-1}$ SHG Fourier amplitudes from n-Si(001) MOS for for $\lambda 
_\omega=725 {\rm \,nm}$ 
($2\hbar \omega =3.43 \,{\rm eV}$). Solid curves are fits to data using the model presented.

\noindent Fig. 9:
Fig. 8. The isotropic SHG component from p-Si(001) MOS structure for $\lambda 
_\omega=730\, 
{\rm nm}$ ($2\hbar \omega =3.41 \,{\rm eV}$) as a function of applied bias. Curves are fits to data 
using the 
model of the DCF screening within "classical" approach for $N_A=1.5 \cdot 10^{15}\, 
{\rm cm}^{-3}$ 
(thin curve) and $N_A=10^{17} \,{\rm cm}^{-3}$ (thick curve) and with surface quantization 
corrections 
(dashed curve).

\noindent Fig. 10:
The normalized four-fold symmetric SHG component $a'=c_4\left( {2\sqrt {2c_8}} 
\right)^{-1}$ 
from p-Si(001) MOS structure for for $\lambda _\omega=730 \,{\rm nm}$ ($2\hbar \omega 
=3.41 \,{\rm eV}$) 
as a function of applied bias. Curves are fits to data using the model of the DCF 
screening within 
"classical" approach for $N_A=1.5 \cdot 10^{15} \,{\rm cm}^{-3}$ (thin curve) and 
$N_A=10^{17}\, 
{\rm cm}^{-3}$ (thick curve) and with surface quantization corrections (dashed curve). Inset: 
voltage 
dependences near zero-point of bias.

\noindent Fig. 11:
The bias dependences of the four-fold symmetric anisotropic SHG component $c_4(U)$ 
for 
several wavelengths of the fundamental radiation and their fit presented by solid lines.

\noindent Fig. 12:
The spectral dependence of module of cubic dipole and quadratic quadruple susceptibility 
around 
the direct two-photon $E_1$ transition extracted from the spectra of the EFISH 
azimuthal 
dependences. Solid lines are fits to data by the Lorenz function with real background. 
Inset: the 
spectral dependence of the efficiency of modulated EFISH for $\Delta U=0.6\, {\rm V}$, 
$\Omega =100\, 
{\rm Hz}$ in p-Si(001) MOS structure.

\noindent Fig. 13:
The experimental static bias dependence of the SHG intensity. The right side shows the 
schematic 
of the low-frequency electromodulation of the SHG signal by the application of the 
superposition 
of DC-bias $U$ and low-frequency squarewave modulation voltage $\Delta U 
(\Omega)$ with the 
amplitude $\Delta U$ and frequency $\Omega$.

\noindent Fig. 14:
The bias dependence of the efficiency of modulated EFISH signal in p-Si(001) MOS 
structure: 
open symbols are experimental dependence of $\alpha (U)$ for for $\Delta U=0.6\, {\rm V}$, 
$\Omega 
=100\, {\rm Hz}$; solid symbols are numerical derivative of the static (DCF-induced) bias 
dependence of 
the EFISH intensity; solid line is the model calculation in accordance of Sections A and 
B in Part 
II.

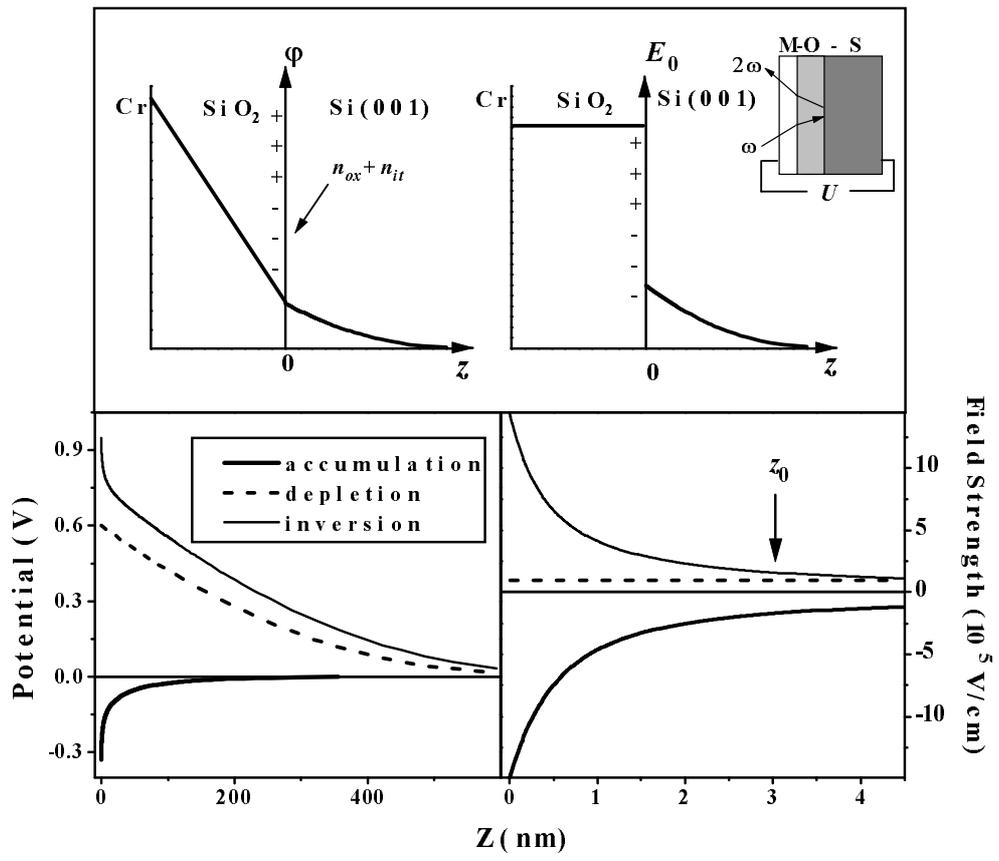
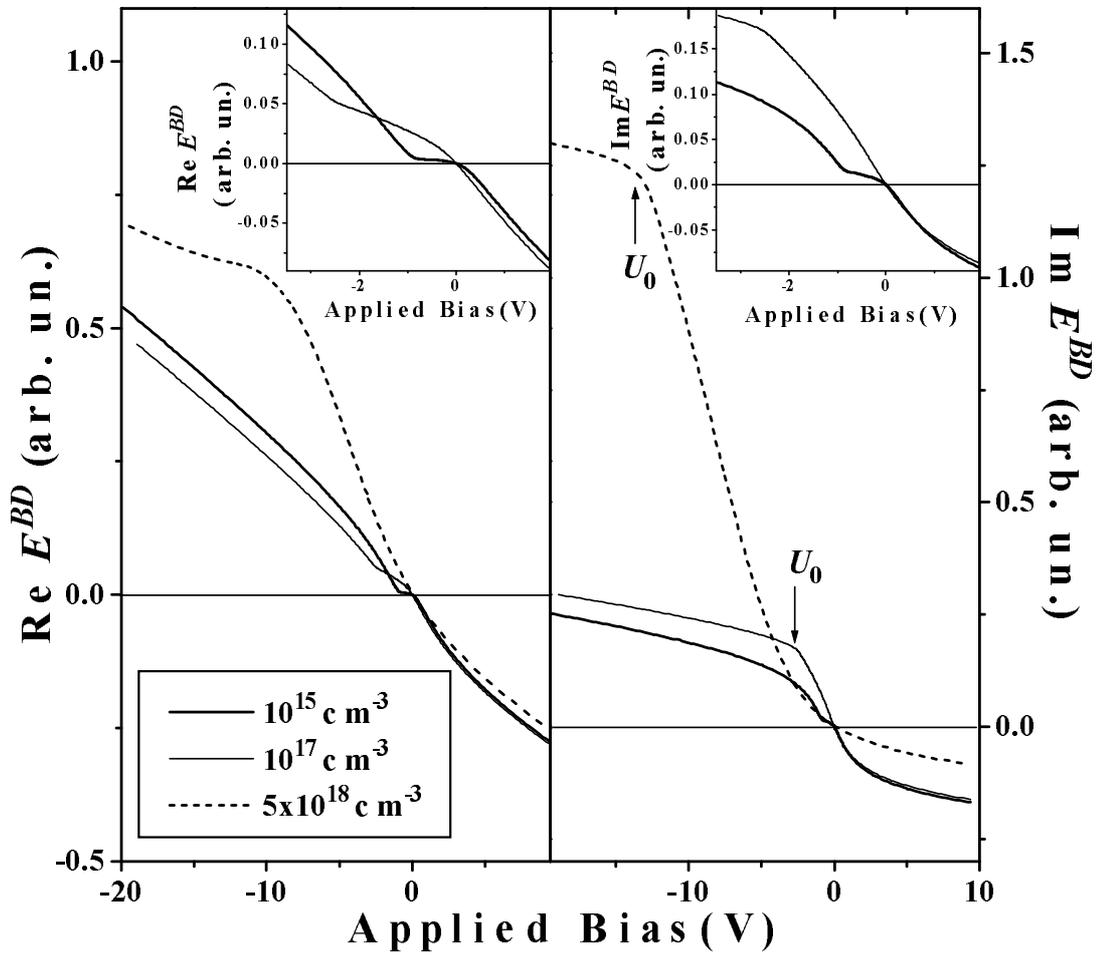
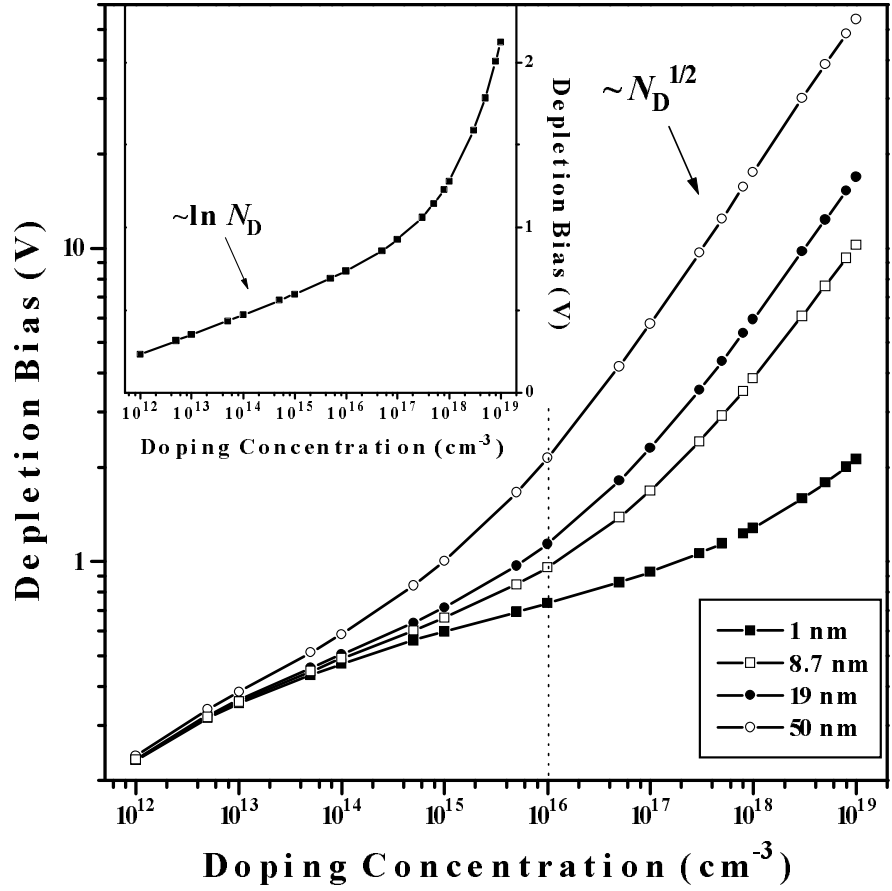
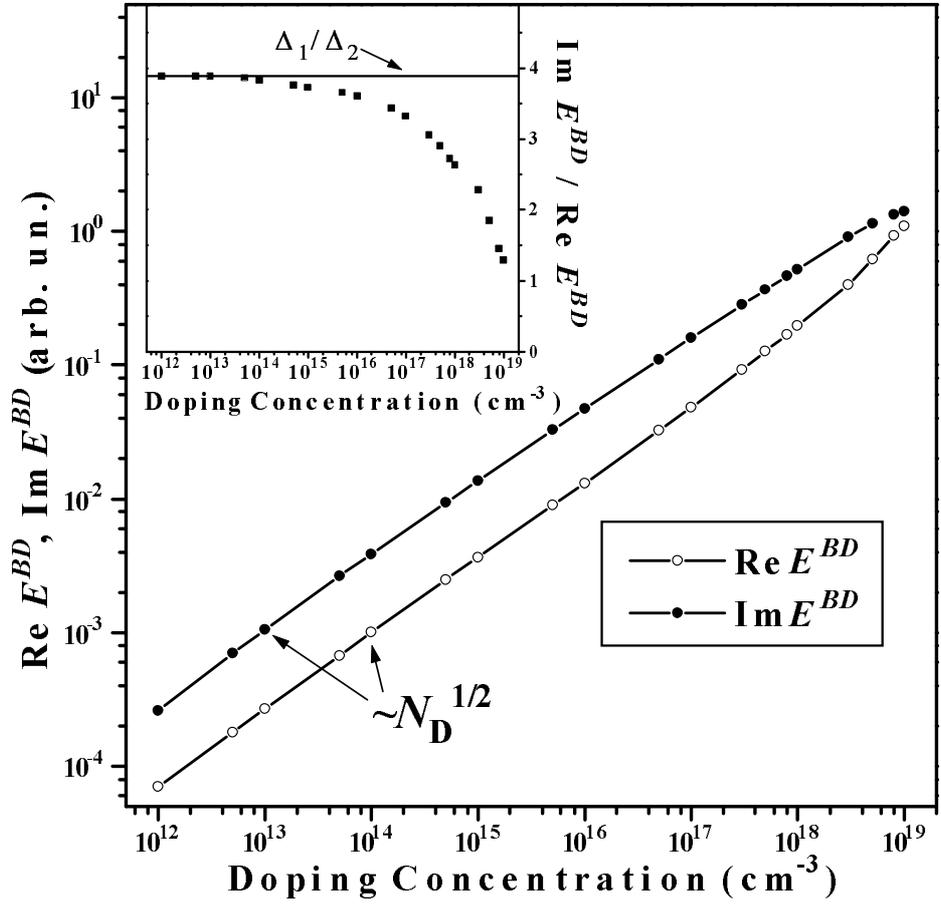
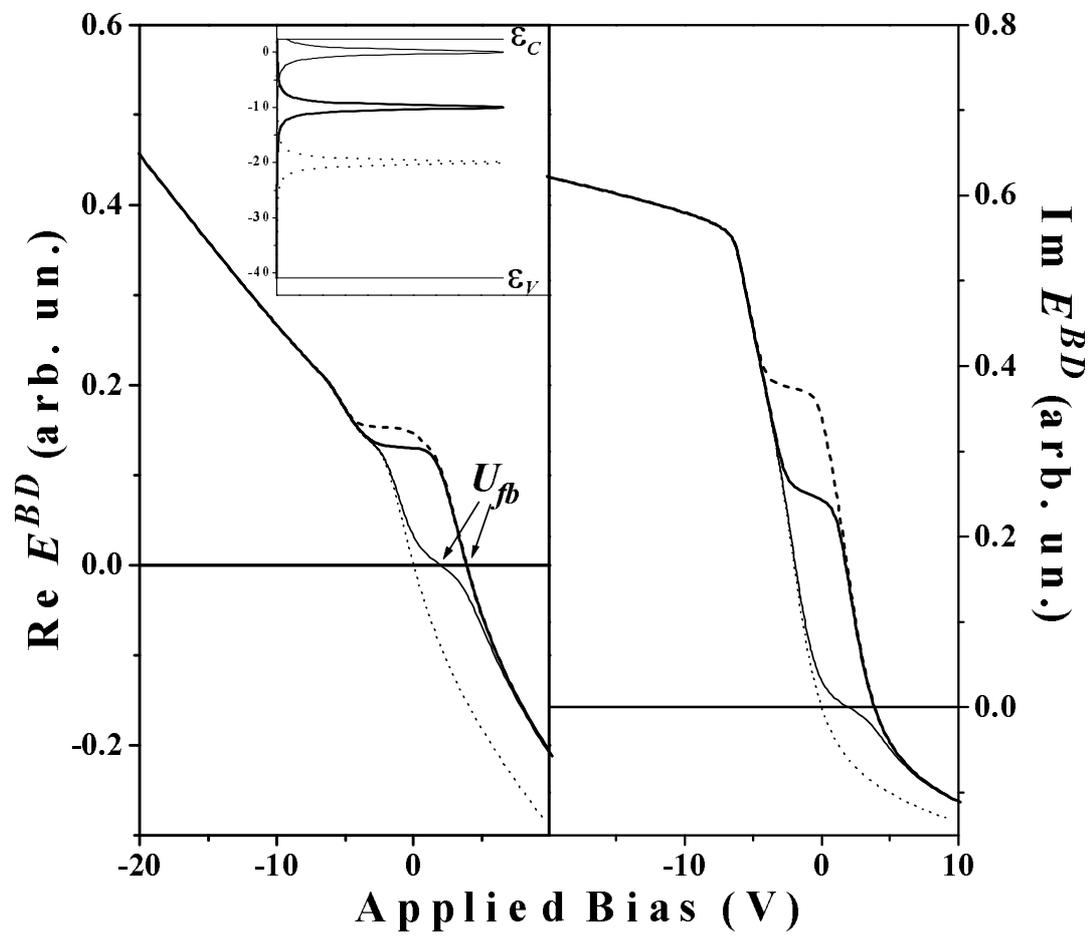
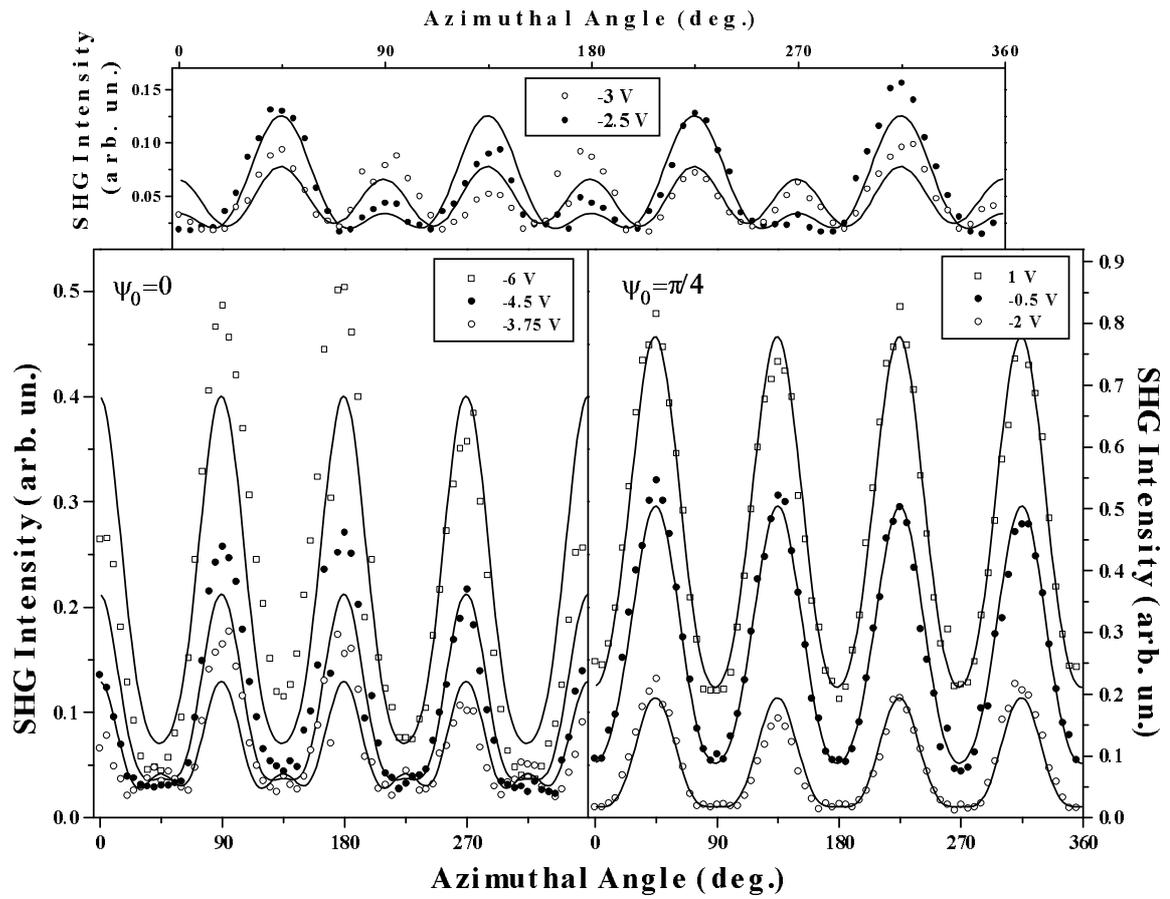
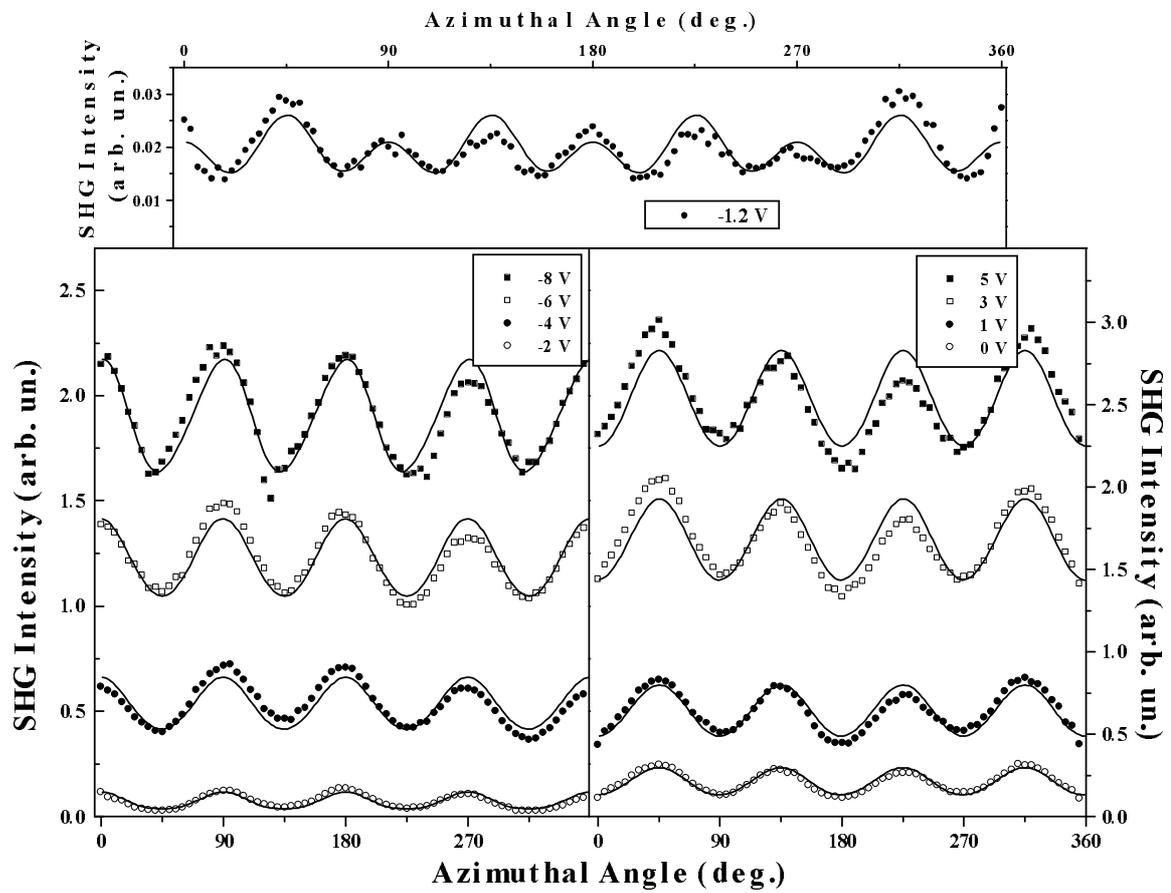
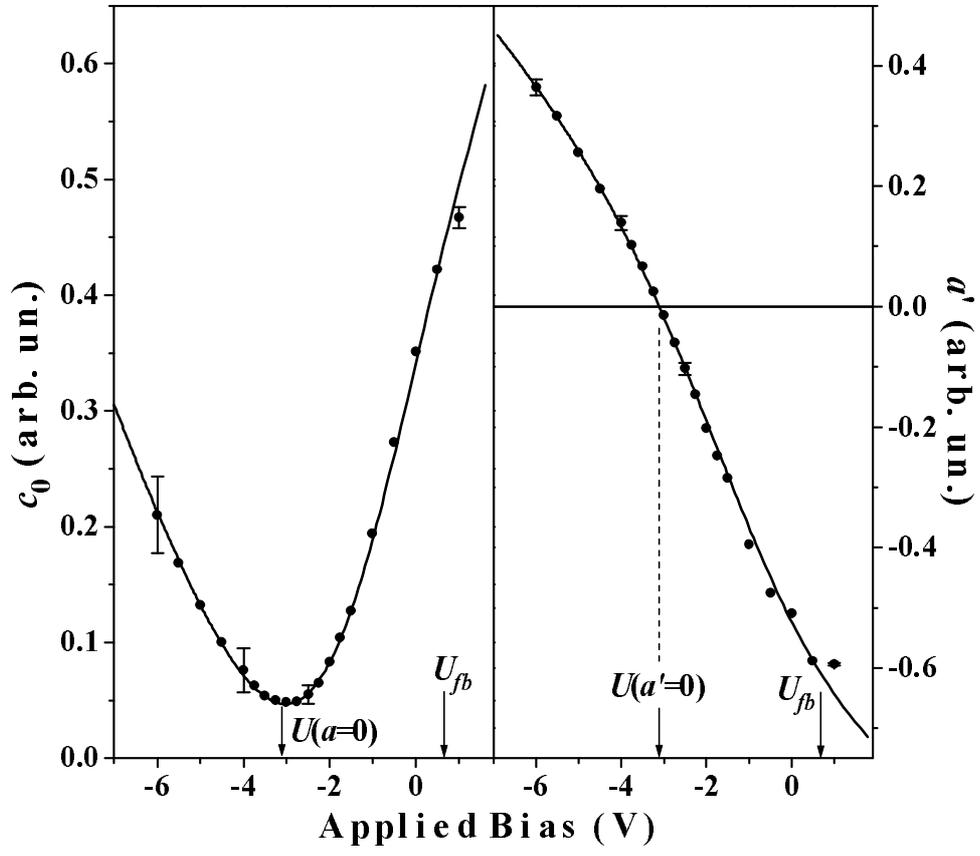
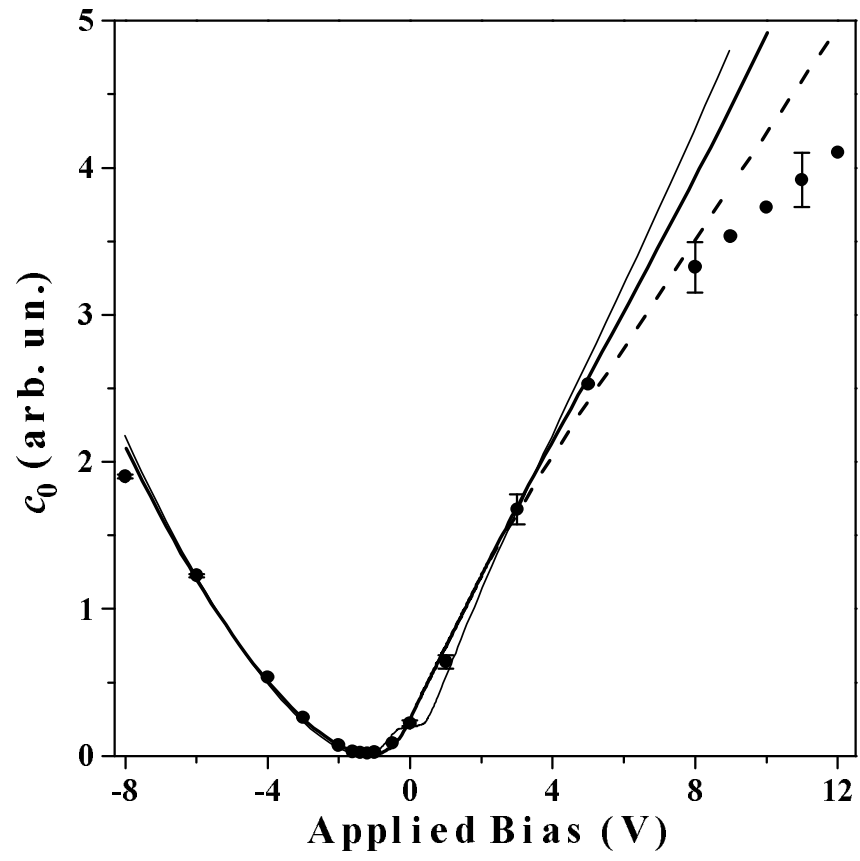
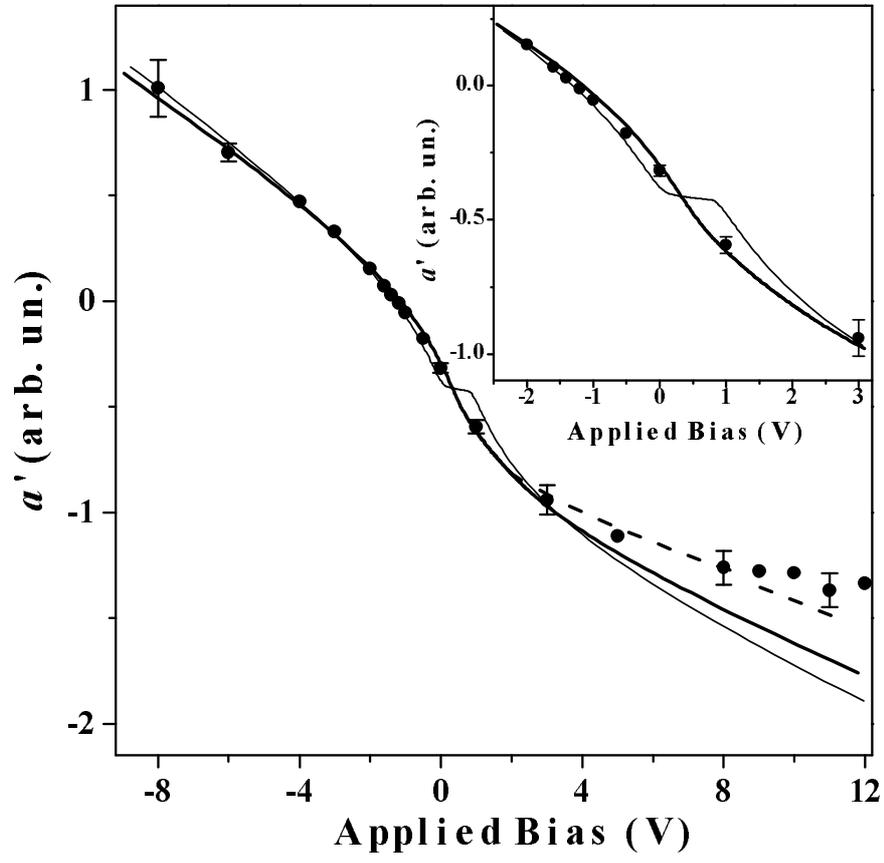
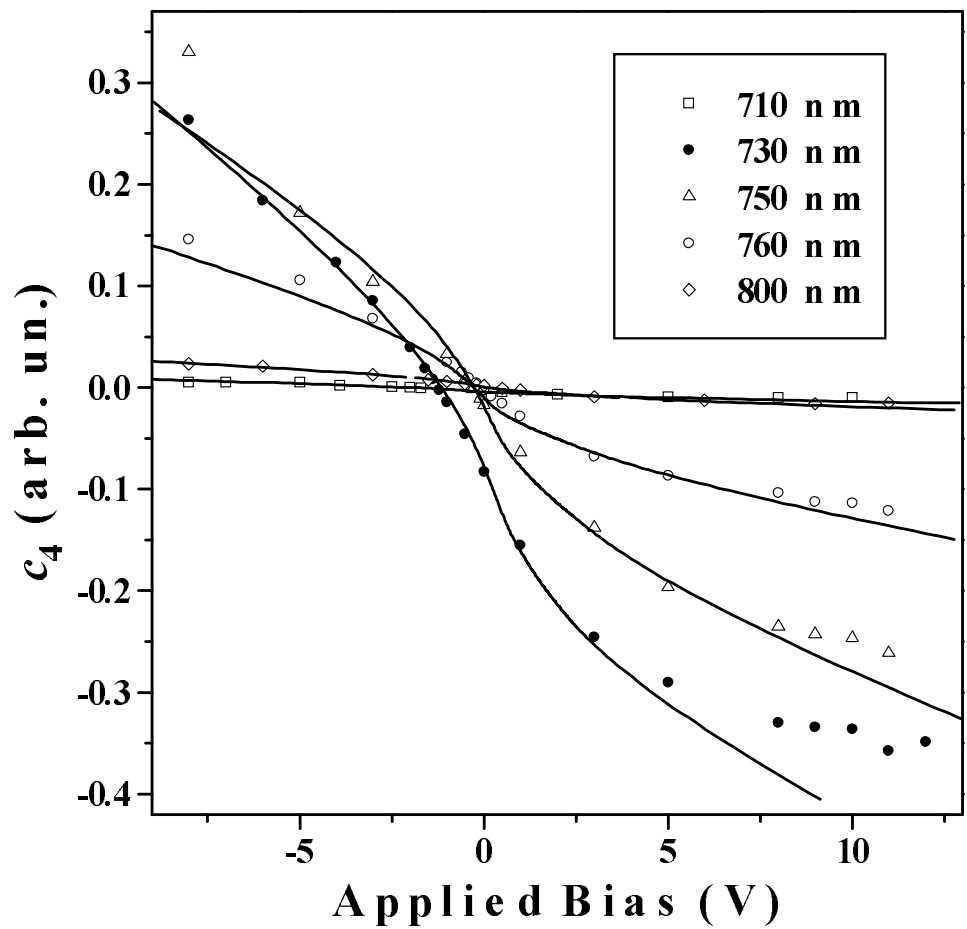
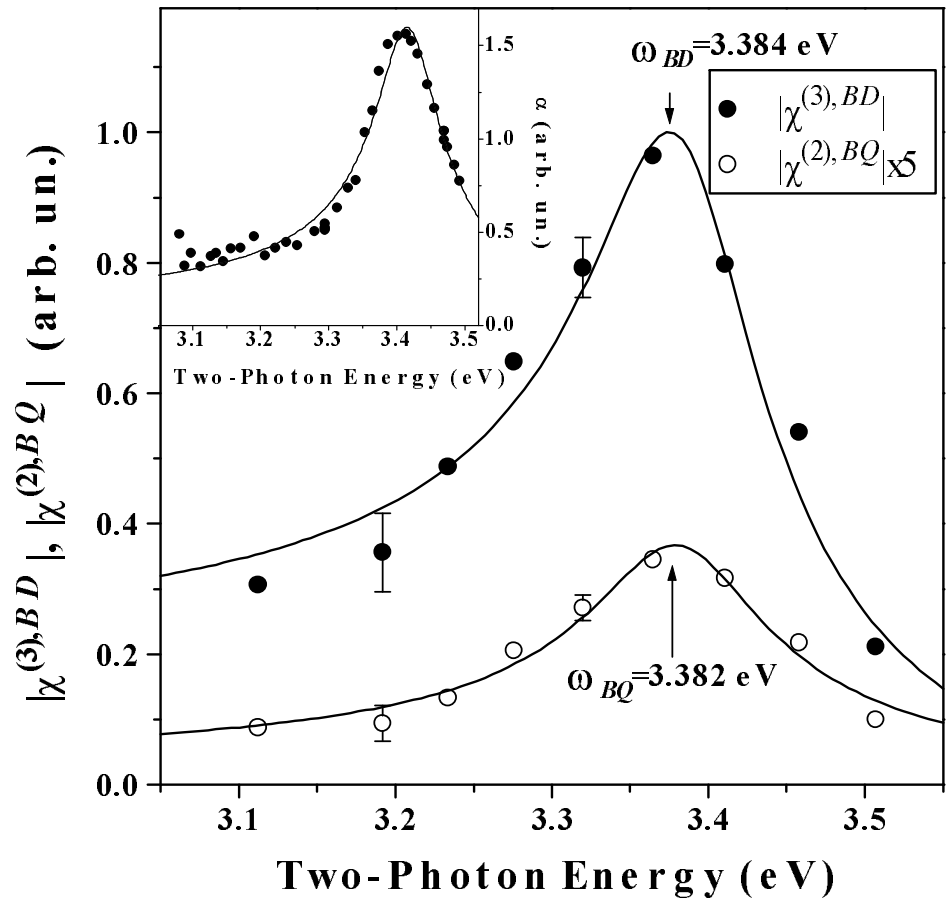
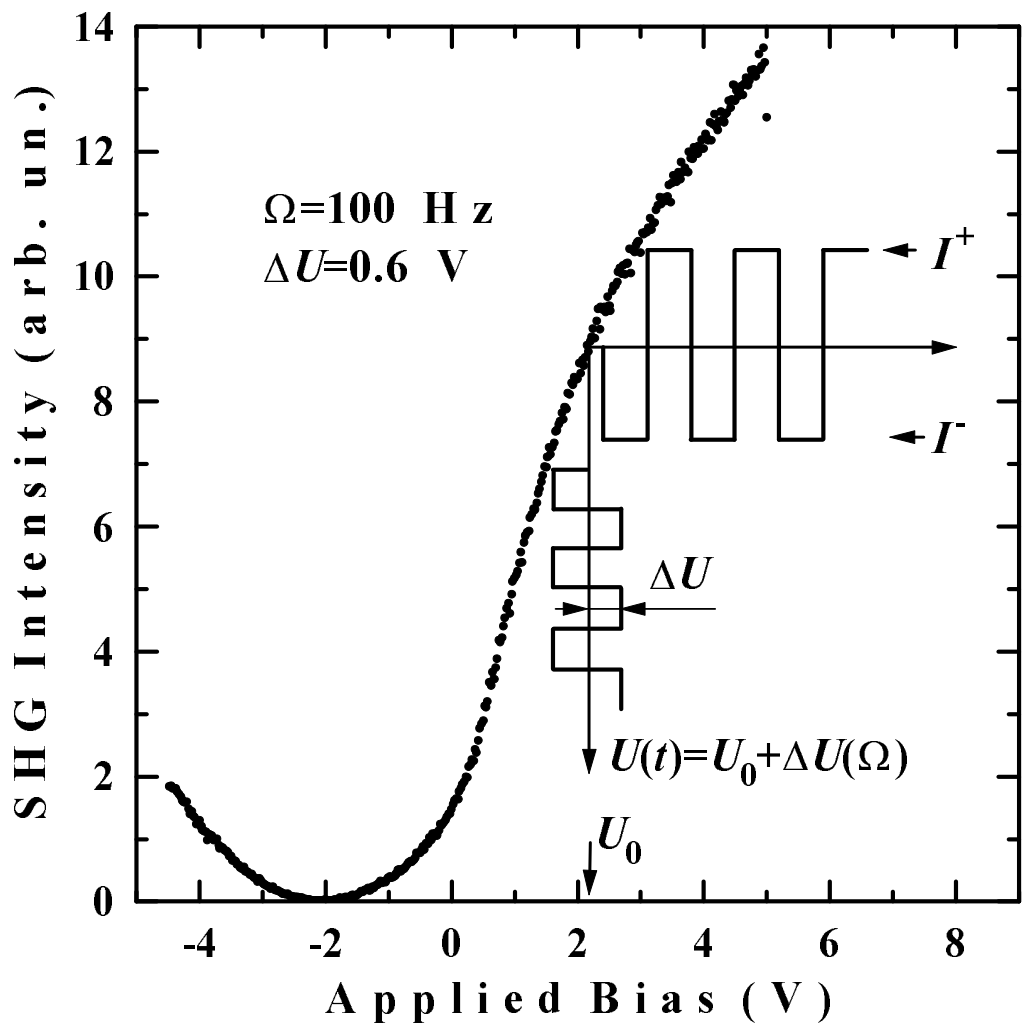
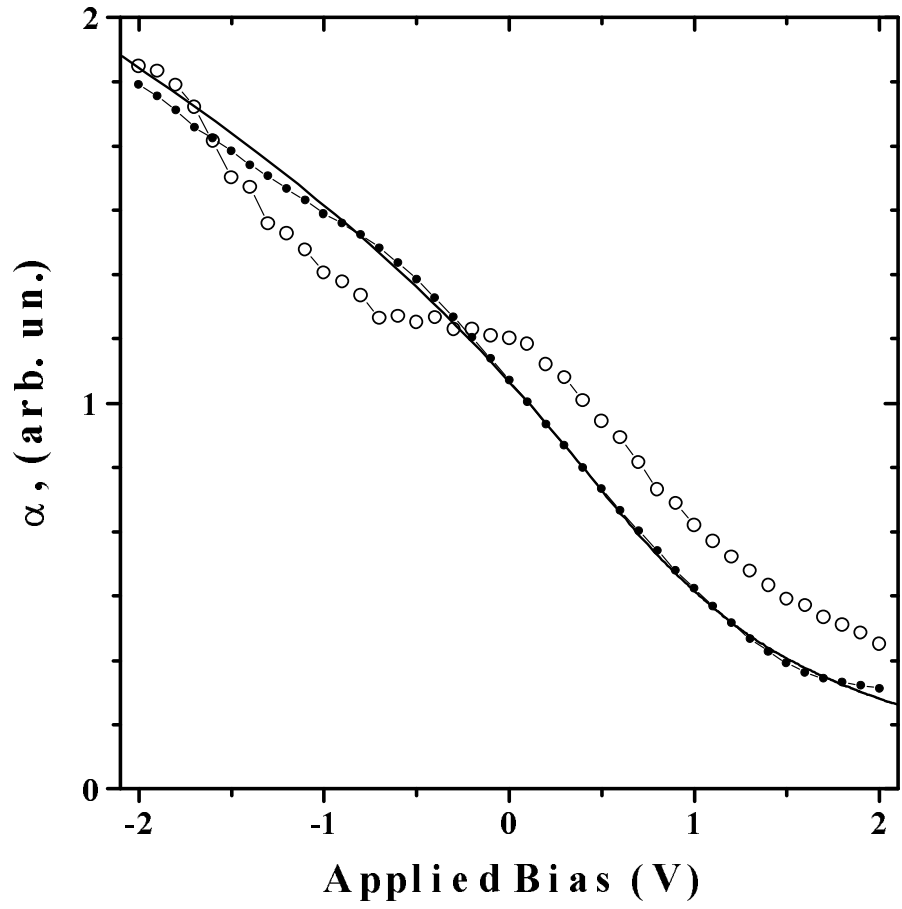
\end{document}